\documentclass[letterpaper,english,aps,pra,amssymb,amsfonts,twocolumn,tightenlines,superscriptaddress,showpacs]{revtex4}
\usepackage[T1]{fontenc}
\usepackage{color}
\usepackage{amsmath}
\usepackage{graphicx}
\makeatletter
\@ifundefined{textcolor}{}
{%
 \definecolor{BLACK}{gray}{0}
 \definecolor{WHITE}{gray}{1}
 \definecolor{RED}{rgb}{1,0,0}
 \definecolor{GREEN}{rgb}{0,1,0}
 \definecolor{BLUE}{rgb}{0,0,1}
 \definecolor{CYAN}{cmyk}{1,0,0,0}
 \definecolor{MAGENTA}{cmyk}{0,1,0,0}
 \definecolor{YELLOW}{cmyk}{0,0,1,0}
 }

\@ifundefined{textcolor}{}
{%
 \definecolor{BLACK}{gray}{0}
 \definecolor{WHITE}{gray}{1}
 \definecolor{RED}{rgb}{1,0,0}
 \definecolor{GREEN}{rgb}{0,1,0}
 \definecolor{BLUE}{rgb}{0,0,1}
 \definecolor{CYAN}{cmyk}{1,0,0,0}
 \definecolor{MAGENTA}{cmyk}{0,1,0,0}
 \definecolor{YELLOW}{cmyk}{0,0,1,0}
 }

\@ifundefined{definecolor}
 {\@ifundefined{definecolor}
 {\@ifundefined{definecolor}
 {\usepackage{color}}{} }{}
}{}

\usepackage[
colorlinks=true,linkcolor=red,citecolor=green,urlcolor=blue,
pdfstartview=FitH,
bookmarksopen=true,bookmarksopenlevel=0,
plainpages=false,pdfpagelabels,
pagebackref=false,
pdftoolbar=false]{hyperref}
\usepackage{nicefrac}

\begin{document}

\title{\mbox{Automated Synthesis of Dynamically Corrected Quantum Gates}}

\author{Kaveh Khodjasteh}
\affiliation{\mbox{Department of Physics and Astronomy, Dartmouth 
College, 6127 Wilder Laboratory, Hanover, NH 03755, USA}}

\author{Hendrik Bluhm} 
\affiliation{2nd Institute of Physics C, RWTH Aachen University, 52074
Aachen, Germany, \\ and JARA, Fundamentals of Future Information
Technologies}

\author{Lorenza Viola}
\email{lorenza.viola@dartmouth.edu}
\affiliation{\mbox{Department of Physics and Astronomy, Dartmouth 
College, 6127 Wilder Laboratory, Hanover, NH 03755, USA}}

\date{\today}
\begin{abstract}
Dynamically corrected gates are extended to non-Markovian open quantum
systems where limitations on the available controls and/or the
presence of control noise make existing analytical approaches
unfeasible.  A computational framework for the synthesis of
dynamically corrected gates is formalized that allows sensitivity
against non-Markovian decoherence and control errors to be
perturbatively minimized via numerical search, resulting in robust
gate implementations. Explicit sequences for achieving universal
high-fidelity control in a singlet-triplet spin qubit subject to realistic
system and control constraint are provided, which simultaneously
cancel to the leading order the dephasing due to non-Markovian
nuclear-bath dynamics and voltage noise affecting the control fields.  
Substantially improved gate fidelities are predicted for current laboratory devices.
\end{abstract}


\date{\today}
\maketitle 

\section{Introduction}

Achieving high-precision control over quantum dynamics in the presence
of decoherence and operational errors is a fundamental goal across
coherence-enabled quantum sciences and technologies.  In particular,
realizing a universal set of quantum gates with sufficiently low error
rate is a prerequisite for fault-tolerant quantum computation
\cite{Knill-Noisy}.  Open-loop control based on time-dependent
modulation of the system dynamics has been extensively explored as a
physical-layer error-control strategy to meet this challenge.  Two
main approaches have been pursued to date: on the one hand, if the
underlying open-system relaxation dynamics is fully known, powerful
variational techniques and/or numerical algorithms from optimal
quantum control theory (OCT) may be invoked to optimize the target
gate fidelity, see {\em e.g.}
\cite{Thomas,Whaley2006,KurizkiBOMEC,Goan2012} for representative
contributions.  On the other hand, dynamically corrected gates (DCGs)
\cite{dcg1} have been introduced having maximum design simplicity and
{\em portability} in mind: close in spirit to well-established
dynamical decoupling techniques for quantum state preservation in
non-Markovian environments \cite{lvdd}, DCG sequences can achieve a
substantially smaller net decoherence error than individual
``primitive'' gates by making minimal reference to the details of the
system and control specifications.  In principle, the use of recursive
control design makes it possible for the final accuracy to be solely
limited by the shortest achievable control time scale
\cite{khodjasteh_arbitrarily_2010}.  Remarkably, simple DCG
constructions underly the fidelity improvement reported for
spin-motional entangling gates in recent trapped-ion experiments
\cite{Hayes}.

While obtaining a detailed quantitative characterization of the noise
mechanisms to overcome is imperative to guarantee truly \emph{optimal}
control performance, this remains practically challenging for many
open quantum systems of interest.  In addition, current approaches for
applying OCT methods to {\em non-Markovian} environments rely on
obtaining suitable simplifications of the open-system equations of
motion (e.g., via identification of a finite-dimensional Markovian
embedding \cite{Thomas} or approximation through time-local coupled
linear equations \cite{Goan2012}) -- which may be technically
challenging and/or involve non-generic assumptions.  Since in DCG
schemes the error cancellation is engineered at the level of the {\em
full} system-plus-environment Hamiltonian evolution, two significant
advantages arise: environment operators may be treated {\em
symbolically}, avoiding the need for an explicit equation of motion
for the reduced dynamics to be derived; in contrast to error-control
approaches designed in terms of gate propagators (notably, fully
compensating composite pulses for systematic control errors
\cite{Levitt,KenReview}), working at the Hamiltonian level allows to
more directly relate to physical error mechanisms and operational
constraints.  Despite incorporating realistic requirements of finite
maximum control rates and amplitudes, analytic DCG constructions
nonetheless rely on the assumption that complete control over the
target system can be afforded through a universal set of {\em
stretchable} control Hamiltonians \cite{dcg1}.  This requirement is
typically too strong for laboratory settings where only a {\em
limited} set of control Hamiltonians can be turned on/off with 
sufficient precision and speed, and universality also relies on 
internal always-on Hamiltonians.  Furthermore, portability
comes at the expenses of longer sequence durations, making DCGs more
vulnerable to uncompensated Markovian decoherence mechanisms.

In this work, we introduce a control methodology that results in an
\emph{automated} recipe for synthesizing DCGs via numerical search.
This is accomplished by relaxing the portability requirement and
utilizing the full details of the control.  While the resulting
``automated DCGs'' (aDCGs) are still synthesized without {\em
quantitative} knowledge of the underlying error sources, they overcome
the restrictive assumptions of analytical schemes and lead to
drastically shorter sequences.  As an additional key advantage, our
Hamiltonian-engineering formulation lends itself naturally to
incorporating robustness against multiple error sources, that can
enter the controlled open-system Hamiltonian in either additive or
multiplicative fashion.  This allows for aDCGs to {\em simultaneously}
cancel non-Markovian decoherence {\em and} control errors, as long as
the combined effects remain perturbatively small.

We quantitatively demonstrate these advantages by focusing on a highly
constrained control scenario -- the two-electron singlet-triplet
(S-T$_0$) spin qubit in GaAs quantum dots (QDs)\cite{Levy2002}.  In
spite of ground-breaking experimental advances
\cite{Coherence,Hendrik,Schulman,Dial}, boosting single-qubit gate
fidelities is imperative for further progress towards scalable quantum
computation and is attracting intense theoretical effort
\cite{Talk,Matthew,DasSarma}. Recently introduced {\sc supcode}
composite-pulse sequences \cite{DasSarma}, for instance, are
(analytically) designed to achieve insensitivity against decoherence
induced by coupling to the surrounding nuclear-spin bath, however they
do not incorporate robustness against voltage noise, which is an
important limitation in experiments \cite{Dial}.  Here, we provide
explicit aDCG sequences for high-fidelity universal control in S-T$_0$
qubits, which cancel the dominant decoherence and exchange-control
errors, while respecting the stringent timing and pulsing constraints
of realistic S-T$_0$ devices.  The resulting sequences use a very {\em
small} number of control variables and a {\em fixed} base pulse
profile, which streamlines their experimental implementation. Up to
two orders of magnitude improvement in gate fidelities are predicted
for parameter regimes appropriate for current experimental conditions.

\section{Control-theoretic setting} 

We consider in general a
$d$-dimensional target quantum system $S$ coupled to an environment
(bath) $B$, whose total Hamiltonian $H$ on ${\cal H}_S \otimes {\cal
H}_B$ reads
\begin{eqnarray}
H & = & [H_S + H_{\text{ctrl},0}(t)]\otimes {\mathbb I}_B + H_e, 
\label{Ham0} \\
H_e & \equiv & H_{e,\text{int}} + H_{e,\text{ctrl}}(t) , 
\nonumber
\end{eqnarray}
where ${\mathbb I}_{B(S)}$ denotes the identity operator on $B$($S$),
$H_S$ accounts for the internal (``drift'') system's evolution in the
absence of control, and the time-dependent $H_{\text{ctrl},0}(t)$
represents the intended control Hamiltonian on $S$.  The total ``error
Hamiltonian'' $H_e$ encompasses the bath Hamiltonian, unwanted
interactions with the bath, as well as deviations of the applied
control Hamiltonian from $H_{\text{ctrl},0}(t)$, subject to the
requirement that the underlying correlation times are sufficiently
long.  Formally, we require that $\Vert H_e\Vert \ll \Vert H(t)\Vert$,
where $\Vert X \Vert$ is the operator norm of $X=X^\dagger$ (maximum
absolute eigenvalue of $X$) \cite{lvdd,dcg1}. In order to ``mark'' the
error sources, we characterize the strength of each independent
contribution to $H_{e,\text{int}}$ in terms of dimensionless
parameters $\{\delta_\alpha\}$, in such a way that, without loss of
generality, we may express
$$ H_{e,\text{int}} = {\mathbb I}_S \otimes H_B + H_{SB} \equiv
\sum_\alpha \delta_\alpha S_\alpha \otimes B_\alpha,$$
\noindent 
with $S_\alpha$ being a Hermitian operator basis on ${\cal H}_S$ and
$B_\alpha$ acting on ${\cal H}_B$, respectively, and the bath internal
Hamiltonian $H_B\equiv \delta_0^e B_0$. We assume that the $B_\alpha$
are norm-bounded but otherwise quantitatively unspecified.  In
particular, if $B_\alpha$ are treated as scalars ($B_\alpha
=\ell_\alpha {\mathbb I}_B$), we may formally recover the limit of
a classical bath, whereby $B_0=0$ and the system Hamiltonian is
effectively modified in a random (yet slowly time-dependent)
fashion. Note that, as long as we are interested in canceling effects
that are first order in the error sources, there is no distinction
between the $B_\alpha$ being actual operators or scalars.  Similarly,
we characterize the independent error sources in
$H_{\text{ctrl},0}(t)$ by letting
$$ H_{e,\text{ctrl}} (t)= \sum_\beta \delta_\beta H_\beta (t) \otimes
{\mathbb I}_B,$$
\noindent 
where $H_\beta (t)$ are known system operators, while the parameters
$\delta_\beta$ remain unspecified.  For notational convenience, we
shall label all the unknown parameters symbolically and collectively
by $\delta \equiv \{\delta_\alpha, \delta_\beta\}$.

In an ideal error-free scenario, $\delta=0$, the system evolves
directly under the action of the control, in the presence of its
internal drift Hamiltonian.  We assume that in this limit, $S$ is {\em
completely controllable}, that is, arbitrary unitary transformations
on $S$ can be synthesized as ``primitive gates'' by suitably designing
$H_{\text{ctrl},0}(t)$ in conjunction with $H_S$.  As mentioned, we
are particularly interested in the situation where the latter is
essential for controllability to be achieved \cite{Remark}.  The
available control resources may be specified by describing
$$H_{\text{ctrl},0}(t)=\sum_a c_a(t) H_a \otimes {\mathbb I}_B,$$ 
\noindent 
in terms of the admissible (nominal) control inputs and Hamiltonians.
Beside restrictions on the set of tunable Hamiltonians $H_a$, limited
``pulse-shaping'' capabilities will typically constrain the control
inputs $c_a(t)$ as system-dependent features of the control hardware.
For concreteness, we assume here that $H_{\text{ctrl},0}(t)$ is
decomposed as a sequence of shape-constrained pulses applied back to
back and also constrain pulse amplitudes $\{ h_i\}$ and durations $\{
\tau_i\}$ to technological limitations such as $h_{\text{min}} \leq
h_i \leq h_{\text{max}}, \tau_{\text{min}} \leq \tau_i \leq
\tau_{\text{max}}$.

Ideally, if the target unitary gate is $Q$, the objective for gate
synthesis is to devise a control Hamiltonian $H_{\text{ctrl},0}(t)$
such that (up to a phase),
\begin{equation}
Q= {\cal T}\hspace*{-0.5mm}\exp\hspace*{-0.5mm}\Big[ -i \int_0^\tau
\hspace*{-1mm} (H_S + H_{\text{ctrl},0}(s)) \,ds \Big],
\label{psynthesis}
\end{equation}
where ${\cal T}$ denotes time ordering and $\tau$ is the running time
of the control.  The ideal evolution naturally defines a
toggling-frame unitary propagator given by
\begin{equation} 
U_Q (t)= {\cal T}\hspace*{-0.5mm}\exp\hspace*{-0.5mm}\Big[ - i
\int_0^t \hspace*{-1mm}(H_S + H_{\text{ctrl},0}(s))\,ds \Big],
\label{tf}
\end{equation}
which traces a path from ${\mathbb I}_S$ to $Q$ over $\tau$.  If $H_e
\ne 0$, application of $H_{\text{ctrl},0}(t)$ over the same time
interval results in a total propagator of the form
$$U^{(\delta)}_{Q[\tau]} \equiv U^{(0)}_{Q[\tau]} \exp(-i
E_{Q[\tau]}^{(\delta)}),$$ 
\noindent 
where $U^{(0)}_{Q[\tau]}= Q$ and $E_{Q[\tau]}^{(\delta)}$ is an
``error action'' operator on ${\cal H}_S \otimes {\cal H}_B$ that
isolates the effects of undesired terms in the evolution \cite{dcg1}:
\begin{equation}
\exp(-i E_{Q[\tau]}^{(\delta)}) = {\cal
T}\hspace*{-0.5mm}\exp\hspace*{-0.5mm}\Big[ -i \int_0^\tau
\hspace*{-1mm} U_Q(s)^\dagger H_e U_Q(s)\, ds \Big].
\label{ea}
\end{equation}
The norm of the error action can be taken to quantify the error
amplitude per gate (EPG) in the presence of $\delta$.  The EPG in turn
upper-bounds the fidelity loss between the ideal and actual evolution
on $S$ once its ``pure-bath'' components are removed.  More
concretely, define
$$\text{mod}_B E\equiv E - \mathbb{I}_S \otimes \mbox{Tr}_S(E)/d , $$ 
\noindent 
that is, a projector that removes the pure-bath terms
in $E$ (note that mod$_B E =E$ if $E$ is a pure-system operator of the form 
$A\otimes {\mathbb I}_B$, as
for a classical bath). Then the following (not tight) upper bound for
the (Uhlman) fidelity loss holds \emph{independently} of the initial
states \cite{khodjasteh_distance,khodjasteh_arbitrarily_2010}:
$$1-f_U \leq \Vert \text{mod}_B E_{Q[\tau]}^{(\delta)} \Vert .$$ 
\noindent 
Thus, reducing the EPG can be used as a proxy for reducing gate
fidelity loss.  While for a primitive gate implementation the EPG
scales linearly with $\delta$, the goal of DCG synthesis is to {\em
perturbatively cancel} the dependence on $\delta$ in
$E_{Q[\tau]}^{(\delta)}$ up to a desired order of accuracy, to realize
the gate in a manner that is as error free as possible as long as
$\delta$ is small.  For simplicity and immediate application, we focus
here on {\em first-order} aDCG constructions, for which $ \Vert
\text{mod}_B E_{Q[\tau]}^{(\delta)} \Vert = O(\delta^{2})$.

\section{Synthesizing Dynamically Corrected Gates }

\subsection{Existential approach} 

Recall that two main requirements are required in first-order
analytical DCG constructions \cite{dcg1,khodjasteh_arbitrarily_2010}:
(i) primitive gate implementations of the generators of a ``decoupling
group'' associated with the algebraic structure of EPGs and (ii)
particular implementations of the target gate $Q$ (as $Q^*$) and the
identity gate (as $I_Q$) as sequences of primitive gates such that
$Q^*$ and $I_Q$ share the {\em same} first-order EPG, making them a
``balance pair''.  While (i) is provided by controllability and leads
directly to a constructive procedure for correcting to the first-order
the identity evolution, (ii) is essential for modifying this procedure
in such a way that the net first-order error cancellation is
maintained, but $Q$ is effected instead.

Generating balance pairs require further adjustment of gate control
parameters to form a {\em controllable relationship} between EPGs of
gate implementations, holding as an identity regardless of the value
of $\delta$ (or $B_\alpha$).  For example, in the absence of drift
dynamics and control errors, such a controllable relationship can be
engineered by ``stretching'' pulse profiles in time while the
amplitudes are reduced proportionally, resulting in different
realizations of the {\em same} target, with EPGs that scale linearly
with the gate duration.  Similarly, in the presence of a
multiplicative control error, primitive gates with physically
equivalent (modulo $2\pi$) angles of rotation result in different EPGs
(note that similar geometric ideas are used in designing composite
pulses \cite{KenReview}).  We argue next that knowing the control
description and marking the error sources $\{\delta_j\}$ does still
lead to (ii) as long as control constraints allow us to tap into a
continuum of different gate implementations.

The multitude of pathways for realizing a primitive gate $Q$ increases
with gate duration/subsegments as a result of availability of more
control choices and ultimately a simpler control landscape
\cite{Moore2011}.  Assume that (A1) such primitive implementations may
be parametrized as $Q[\tau]$.  We aim to show that a balance pair or,
alternatively, a direct cancellation of the EPG of $Q$ may be found.
The gist of our argument is most easily given for a single qubit, with
the Pauli operators chosen as the operator basis $\{S_\alpha\}$ for
error expansion. Using the fact that the interactions among different
error sources can be ignored up to the first order, the basic idea is
to start with the first error source, $\delta_1$, and then use the
resulting gates recursively for the next error source until all error
sources are exhausted.  Ignoring error sources other than $\delta_1$,
let us thus expand $E^{(\delta_1)}_{Q[\tau]} = \delta_1 \sum_\alpha
e_{Q[\tau],\alpha} S_\alpha \otimes B_1$.

Assume in addition that (A2), as a function of the parameter $\tau$,
the range of the real-valued functions $e_{Q[\tau],\alpha}$ extends to
infinity in positive or negative directions.  Consider now
``projection blocks'' composed of two Pauli gates applied back to
back, that is,
$$P_\alpha [\tau_\alpha]\equiv S_\alpha [\tau_\alpha] S_\alpha
[\tau_\alpha],$$ 
\noindent 
with a corresponding EPG given by $2\delta_1 e_{S_\alpha
[\tau_\alpha],\alpha}S_\alpha\otimes B_1$, which is {\em purely along
$S_\alpha$}.  By virtue of (a2), we can find a continuum of
$(\tau,\tau_\alpha)$ pairs such that $2 e_{S_\alpha
[\tau_\alpha],\alpha} =\pm e_{Q[\tau],\alpha}$ for all Pauli
directions $\alpha$, meaning that we may reproduce each error
component in $E^{(\delta_1)}_{Q[\tau]}$ {\em up to a sign}. Those
Pauli components $\alpha_{-}$ that reproduce error with a negative
sign are combined in sequence with $Q[\tau]$ to form a longer gate
$$Q^{*}=Q[\tau] \prod_{\alpha_{-}}\!P_{\alpha_{-}}
[\tau_{\alpha_{-}}].$$ If all Pauli components can be matched with
negative signs, the resulting gate will cancel all error components
and a DCG construction is provided by $Q^*$. Otherwise, the Pauli
components ${\alpha_{+}}$ that are matched with positive sign, are
combined to produce an identity gate,
$$I_Q=\prod_{\alpha_{+}}\!P_{\alpha_{+}} [\tau_{\alpha_{+}}],$$ that
matches the error of $Q^{*}$. Hence, $(Q^*,I_Q)$ form a balance pair
and can be used to produce a continuum of constructions of a DCG gate
$Q^{(1)}[\tau]$ that cancel the error source $\delta_1$. Provided that
the assumption (A2) remains valid for this new composite
constructions, we can repeat the procedure to remove the other error
sources.

We remark that assumption (A2) essentially implies that the domain of
the errors as a function of implementation parameters for a fixed
unitary gate is not compact, so that arbitrary magnitudes of each error component 
can be sampled by appropriately choosing the implementation parameters. 
Such arbitrary large domains need not not exist in the primitive gate implementations 
(naturally or due to constraints), or only discrete error values may be reachable.  
Nonetheless, we may still enlarge the accessible range of errors for 
the target gate $Q$ by attaching a continuously parametrized family of 
identity gates. Universal controllability of the system implies that not only any 
gate $U$ but also its inverse $U^{-1}$ may be reached. 
Implementing $U$, followed by its inverse $U^{-1}$, produces an 
implementation of the identity $I_{U}$ ``parametrized'' by the original gate 
$U$. In the absence of degeneracies (relationships between the errors that 
could be used separately to provide a balance pair), the EPG 
associated with $E_{U}$ has then a continuos domain. Clearly, applying $I_{U}$ 
followed by the target gate $Q$ still realizes the gate $Q$ but the resulting
EPG is now given by $E_{Q}+E_{I_{U}}$, which is parametrized by $U$. By 
applying sufficiently many copies of $I_{U}$ before applying $Q$, the 
error can be extended to arbitrary large domains as desired.

\subsection{aDCGs: Computational Approach} 

The aDCG sequences generated by following the above existential 
argument tend to be far too long and complex to be useful in realistic control
scenarios. Also note that in principle, the construction of single- or
two-qubit DCGs in $n$-qubit registers may be handled similarly by
using multi-qubit Pauli operators as a basis for EPG expansion and for
building projection blocks. However, the sequence complexity tends in
this case to also grow exponentially with $n$
\cite{khodjasteh_arbitrarily_2010}, making the need for more efficient
synthesis procedures even more essential.  Just as complete
controllability provides an existential foundation to numerical OCT
approaches for unitary gate synthesis when $H_e=0$ \cite{Matthew}, our
argument legitimates a numerical search for aDCGs in the presence of
$H_e$.  Since the objectives of gate realization and perturbative
error cancellation are not inherently competing, the numerical search
can be described as {\em multi-objective minimization} problem, as we
detail next.

Let the nominal control Hamiltonian $H_{\text{ctrl},0}(t)$ be
parametrized in terms of control variables $\{x_{i}\}$ and define
objective functions as follows:
\begin{eqnarray}
F(\{x_{i}\}) &= & 
\text{dist}( U_{Q[\tau]}^{(\delta=0)}, Q ), \label{o1} \\
%=[1-\vert\text{Tr}(Q^{\dagger} U_{Q[\tau]}^{(\delta=0)} )\vert/d]^{1/2}, \\
B_j G_j(\{x_{i}\}) & = & \Vert \partial \,\text{mod}_B
E_{Q[\tau]}^{(\delta)} / \partial \delta_j \Vert_{\delta=0}
\label{o2},
\end{eqnarray}
where $j$ labels independent error sources and $B_j$
\emph{symbolically} denotes bath operators that mark error sources in
$H_{e,\text{int}}$ (recall that $B_j={\mathbb I}_B$ for control error
sources) to ensure that $F$ and $G_j$ only depend on the known
quantities $x_i$.  Minimizing only the first objective, $F=0$,
corresponds to achieving exact ideal primitive gate synthesis,
Eq. (\ref{psynthesis}).  As an appropriate distance measure for
unitary operators in Eq. (\ref{o1}), we use
\begin{equation}
\text{dist}( U, V ) =[1-\vert\text{Tr}(U^{\dagger}
V)\vert/d]^{1/2},
\label{dist} 
\end{equation}
which is a standard phase-invariant choice \cite{Matthew}.  Minimizing
the objectives in Eq. (\ref{o2}) corresponds to first-order
sensitivity minimization.  Thus, solving for $F=0=G_j, \forall j$,
results in an implementation of $Q$ that is {\em insensitive to the
perturbative parameters} $\delta_{j}$, yielding a {\em robust control
solution} as long as $\delta_j$ is small.

Evaluating $G_j$ apparently requires solving the full time-dependent
system-plus-bath Schr\"odinger equation parametrized by the
controllable pulse shapes. In fact, once the error sources (including
bath operators) are treated as first-order {\em symbolic variables},
$G_j$ can be evaluated by effectively solving the Schr\"odinger
equation on the system only, in order to determine the appropriate
toggling-frame propagator, Eq. (\ref{tf}), and then evaluate the
required error action, Eq. (\ref{ea}), by invoking a Magnus expansion
\cite{dcg1}.  Specifically, if the control variables $x_i \equiv
\{(\tau_{i},h_{i})\}$, the sequence propagator reads
$$U_{Q[\tau]}^{(\delta)} \equiv U_{x_{n}}^{(\delta)}(\tau_n)\cdots
U_{x_{1}}^{(\delta)}(\tau_1), $$ 
\noindent 
where $\tau = \sum_i \tau_i$ and $U_{x_{i}}^{(\delta)}(s)$ is the
$i$-th pulse propagator corresponding to the variable $x_{i}$ at time
$s$, with its associated first-order error action,
$$E_{U_{x_{i}}}^{(\delta)} = \int_{0}^{\tau_i}
U_{x_{i}}^{(0)}(s)^{\dagger} H_{e} U_{x_{i}}^{(0)}(s)ds.$$ 
\noindent 
To the first order in $\delta$, the total EPG is in turn given by 
$$E_{Q[\tau]}^{(\delta)}=\sum_i
V_{i}^{\dagger}E_{U_{x_{i}}}^{(\delta)} V_{i},$$
\noindent 
where $V_{i}\equiv U_{x_{i-1}}^{(0)}\cdots U_{x_{1}}^{(0)}$ denote the
``partial'' product of gate propagators up to and excluding the
$i$-the gate \cite{dcg1}.  While, as noted, for a first-order aDCG the
resulting accuracy $\text{mod}_B E_{Q[\tau]}^{(\delta)} \Vert
=O(\delta^2)$, a more quantitative estimate of the actual conditions
of applicability requires estimating the dominant uncorrected
second-order errors.  Technically, this can be carried out by means of
standard algebraic techniques however is not straightforward
\cite{serre} and beyond our present scope.  Instead, we focus in what
follows on addressing the construction and performance of first-order
aDCGs in concrete illustrative settings.

\section{Application to singlet-triplet qubits} 

Consider first the following single-qubit specialization of
Eq. (\ref{Ham0}):
\begin{eqnarray}
H & = & \frac{1}{2}\Big[ {B} \sigma_x + J_0(t)\sigma_z \Big]\otimes
{\mathbb I}_B +H_e ,
\label{Ham1} \\
H_e & = & {\mathbb I}_S \otimes \delta_0 B_0 + 
\sigma_x\otimes \delta_1 B_x +
J_0(t)\delta_2 \sigma_z, \nonumber
\end{eqnarray}
where the operator-valued $\delta_1 B_x$ and the system drift $B$
couple to the system along $\sigma_x$ and the nominal control $J_0(t)$
and a multiplicative error $\delta_2$ couple along $\sigma_z$.
Although explicitly included, the bath internal Hamiltonian does not
play a role in the first-order removal of decoherence and is
automatically accounted for in mod$_B$.  On the other hand, the drift
term $B$ is essential for complete controllability and analytical DCG
constructions are not viable even in the limit $\delta_2 \rightarrow
0$.  Thus, the need to effectively address {\em both} noise sources
$\delta_1, \delta_2$ for a {\em generic} operating point $B$ mandates
the use of aDCGs.

While useful as a template for single-axis control scenarios in the
presence of internal drift and dephasing, a semi-classical version the
above model is relevant, in particular, to describe a universally
controllable S-T$_0$ qubit.  In this case, the logical qubit subspace
is spanned by $\{ |S\rangle , |T_0\rangle \}$, the singlet and triplet
states of two electrons on separate QDs
\cite{taylor,Hendrik,Schulman,DasSarma} and, provided that the number
of bath nuclear spins is sufficiently large \cite{taylor}, the
following simpler Hamiltonian is appropriate and widely used
for this system \cite{Matthew,DasSarma}:
\begin{eqnarray}
H = \frac{1}{2}\Big[{B + \delta B(t)} \Big]\sigma_x +
\frac{1}{2}{J(t)} \sigma_z.
\label{STq}
\end{eqnarray}
Physically, the drift term $B$ is a known static magnetic field
gradient between the two QDs that includes an Overhauser field from
the nuclear spin bath, $\delta B(t)$ (corresponding to $2\delta_1
B_x$) accounts for random fluctuations of $B$ due to coupling to
nuclear flip-flop processes \cite{DasSarma2009}, and $J(t)$ is the
exchange splitting. In practice, $J(t)$ is tuned by control of an
electrostatic gate voltage \cite{Hendrik}, and voltage fluctuations
due to charge noise result in a noisy control Hamiltonian of the form
$J(t)=J_{0}(t)(1+\delta J(t))$, where $\delta J(t)$ thus corresponds
to $2\delta_2$.  We assume that both noise sources may be treated as
Gaussian quasi-static processes, with their ``run-to-run''
distribution being characterized by standard deviations
$\sigma_{\delta B}$ and $\sigma_{\delta J}$.  While in practice the
noise is not completely static, we expect our considerations to remain
valid as long as high-frequency noise components decay sufficiently
fast and the resulting aDCGs are short relative to time scales over
which white charge noise may become important.  Phenomenologically,
the dephasing induced by the fluctuating Overhauser field 
is consistent with a power-law noise spectrum of the form 
$S(\omega)\sim \omega^{-2}$ over a wide spectral range 
\cite{Mike}.  Likewise, recent experiments indicate that 
voltage noise also arises overwhelmingly due to low-frequency 
components with an approximate $1/f$ decay 
at low operating temperatures \cite{Dial}.  From
experimentally measured values of $T_{2}^{\star}$, we use here
$\sigma_{\delta B/(2\pi)} \lesssim 0.15$ MHz \cite{Coherence} and
$\sigma_{\delta J}\lesssim 1/50$ \cite{jnoise}.

In constructing aDCGs, we shall choose values of the internal drift
($B$) and of the nominal control field ($J_0(t)$) appropriate for the
QD setting of Eq. (\ref{STq}).  We stress, however, that the {\em
same} solution is found from (and applies to) the fully quantum model
Hamiltonian of Eq. (\ref{Ham1}).  In practice, the drift term $B$ can
be set to a fixed value $B/2\pi \in [0.03,0.2]\,$GHz, which we choose
at $0.1\,$GHz \cite{boptim}.  The control field $J_0(t)/2\pi$ is taken
to be positive and smaller than $J_{\text{max}}=0.3\,$GHz. We
recognize the finite rise, delay, and drop times associated with pulse
generators by \emph{fixing} a pulse profile. Thus, during each pulse,
with time $t'$ measured from the pulse start, the control function
$J_0(t')$ is given by $h_i c(t'/\tau_i)$, where $c(x)$ is the pulse
shape function. We digitize the pulse shape function for numerical
evaluation.  In contrast to merely bounding the pulse times and
allowing pulse durations as extra control variables, we enforce the pulse
durations to be {\em fixed} at $\tau_i\equiv \tau = 3\,$ns, that is
compatible with the currently most widespread pulse generators
temporal resolution of $0.83\,$ns.  The search space of the pulse
amplitude control variables is thus given by $x_i=h_i$.  While 
removing $\tau_i$ from the control variables results in more severe 
constraints, it also corresponds to a reduction of the search space. We 
verified that all of our results were reproduced with variable but 
lower-bounded pulse widths as well.
The objective functions $F$, $G_1$ and $G_2$ are computed explicitly in
terms of each constituting pulse parameter $h_i$ according to the
general procedure described in Sec. III.B.

\begin{figure}[t]
\begin{centering}
\includegraphics[width=0.9\columnwidth]{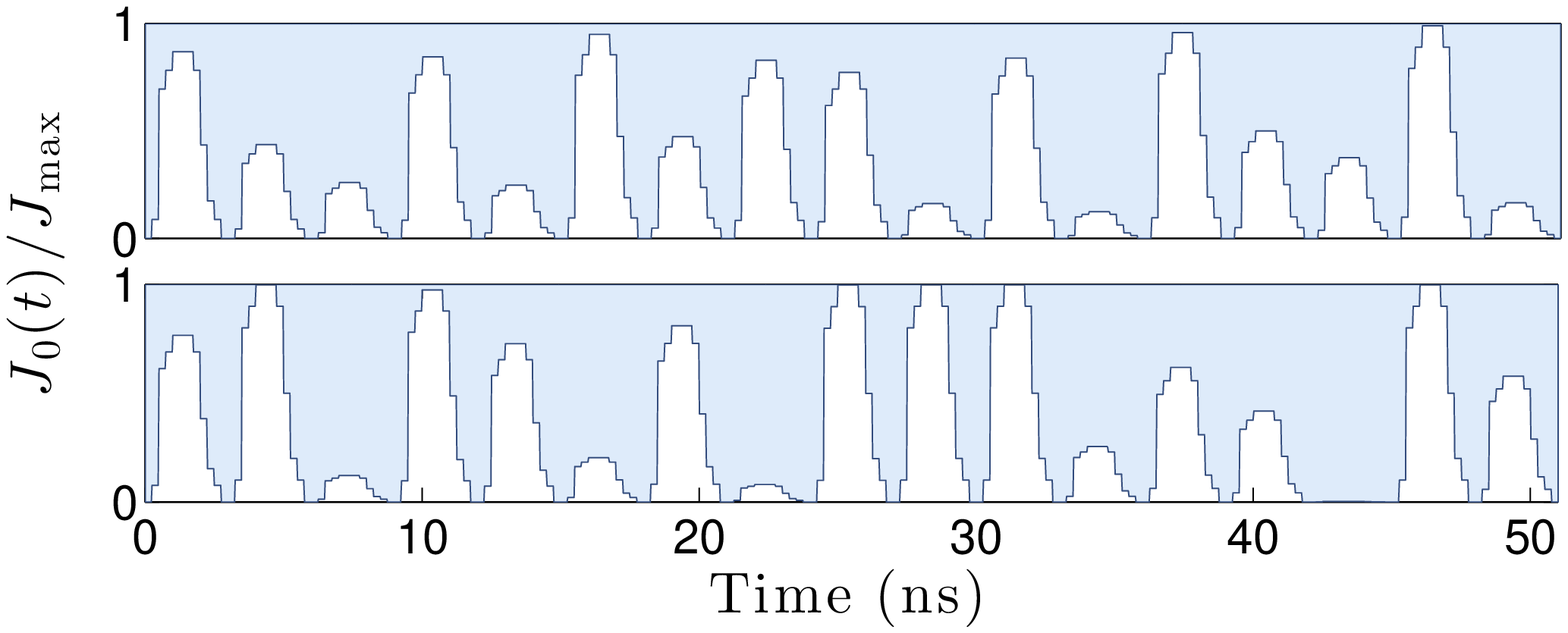} \\
\includegraphics[width=0.9\columnwidth]{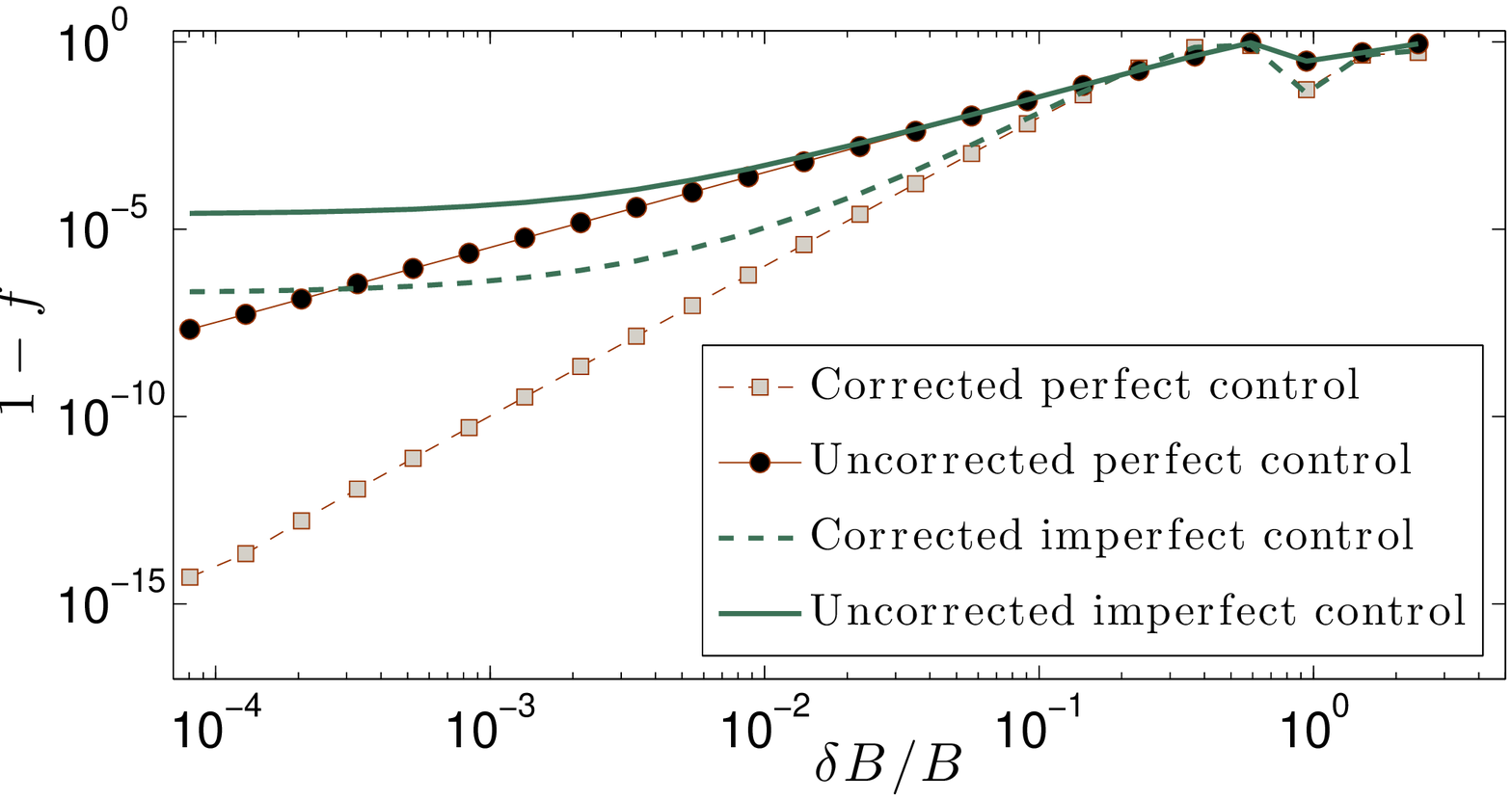}
\par\end{centering}
\vspace*{-1mm}
\caption{(Color online) Upper panel: aDCG control profiles for $\exp(-i\pi
\sigma_z/8)$ (top) and the Hadamard gate (bottom).  Lower panel: 
Fidelity loss for uncorrected vs. corrected gates, evaluated as 1$-f =
\text{dist}(U_{Q}^{(\delta B , \delta J)}, Q)^2$ 
[Eq. (\ref{dist})], for $Q=\exp(-i\pi \sigma_z/8)$ as a function of
relative magnetic field gradient error $\delta B/B$ 
($B/(2\pi)= 0.1\,$GHz).   $\delta J=0$ corresponds to 
perfect exchange control, whilst $\delta J=0.01$ is close to current 
realistic levels.  The fact that $\delta J=0.01$ is {\em fixed} is
responsible for the eventual performance plateaux where the latter dominates 
over $\delta B$ effects.  Nevertheless, the aDCG advantage
{\em is always maintained}.   \vspace*{2mm}}
\label{fig:bars} 
\end{figure}

For the resulting multi-objective minimization, we introduce numerical
weight factors $\lambda_{1}$ and $\lambda_{2}$ and form a single
objective function $$O(\{h_{i}\})\equiv F +\lambda_{1}G_1+
\lambda_{2}G_2.$$  Choosing small values of $\lambda_i$ ($=10^{-5}$)
work best in directing the search from solutions that synthesized the
target gate only at ($F=0$) first, and then towards the
error-corrected solution $F=0, G_1=0=G_2$.  Motivated by our existential
argument, the intuition is to avoid the local minima associated with
multiple objectives and focus on a single objective which, once
realized, will give weight to the other objectives iteratively. We
solve each aDCG search problem using off-the-shelf (Matlab's
\textsc{fmincon} function) search routine (within minutes), with the 
default choice for solving constrained nonlinear optimization without
specifying a precalculated gradient or Hessian.  We start the search
with a small number of pulses, $n$, which is then incremented until
the minimal value of the objective function comes close to the machine
precision ($\approx 10^{-16}$). Fig. \ref{fig:bars} (top) depicts the
synthesized control profiles for a universal set of single-qubit
aDCGs.

\begin{figure}[t]
\begin{centering}
\includegraphics[width=0.9\columnwidth]{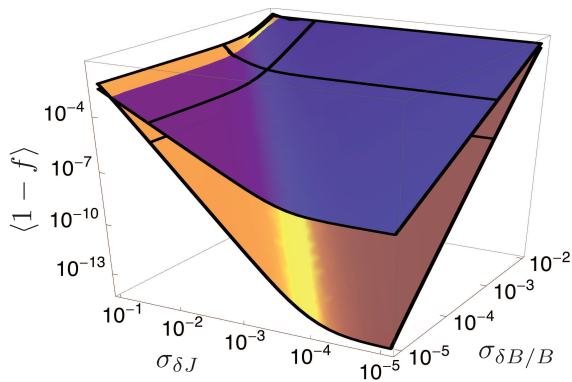}
\par\end{centering}
\vspace*{-1mm}
\caption{(Color online) Ensemble averaged fidelity loss as a function
of relative magnetic field-gradient and exchange-control noise for uncorrected
(top, dark) vs. corrected (bottom, light) implementation of
$Q=\exp(-i\pi \sigma_z/8)$.  The solid (black) lines on each surface 
correspond to typical values $\sigma_{\delta J} \approx 0.02$ and $\sigma_{\delta B/B}
\approx 10^{-3}$ for current S-T$_0$ devices.
\label{fig:3d}}
\end{figure}

Once aDCG sequences are found, evaluating their effectiveness for the
S-T$_0$ qubit can take direct advantage of the effectively
closed-system nature of the model Hamiltonian in Eq. (\ref{STq}), thus
avoiding the need of explicit spin-bath simulations and quantum process 
tomography.
Fig. \ref{fig:bars} (bottom) depicts the fidelity loss for an
uncorrected ($n=3$ pulses, obtained through the same numerical
procedure with $\lambda_1=\lambda_2=0$) vs. corrected implementation
($n=17$ pulses).  The higher slope of the fidelity loss as a function
of $\delta B$ when $\delta J=0$ is the signature of a perturbative
error cancellation and the aDCG advantage is maintained even with
$\delta J>0$, implying robustness with respect to both error types.

In order to make contact with experimentally relevant
ensemble-averaged fidelities, we further evaluate the average of
single-run fidelities $f(\delta B,\delta J)$ with respect to noise
realizations, by assuming that $\delta B$ and $\delta J$ are
independent and normally distributed random variables with variance
$\sigma^2_{\delta B}$ and $\sigma^2_{\delta J}$. The results are
summarized in Fig. \ref{fig:3d}.  Both noise sources adversely impact
the expected gate fidelity but aDCGs are far less affected, resulting
in robust gates roughly as long as $\sigma_{\delta B/B}+\sigma_{\delta
J}\lessapprox 0.1$.

\section{Conclusion} 

Our procedure can be interpreted as a
\emph{automated gate compiler} which incorporates detailed information
about the controllable parameters and their range of operations, along
with qualitative information about the error sources affecting the
evolution.  Compared to mere (primitive) gate synthesis, the resulting 
increase in complexity scales proportionally to the number of error sources.  
Our approach applies to any Hamiltonian control setting, and
for weak enough error sources, even higher-order cancellation can
be achieved in principle.  

Thanks to the slow dynamics of the nuclear spin bath and fast control
pulses available, electron spin qubits provide an ideal experimental
testbed for validating our approach.  While additional experimental
details may be captured in more sophisticated ways, we believe that
our framework is general and flexible enough for its effectiveness not
to be compromised.  In particular, further analysis is needed to
quantify the effect of white electrical noise on aDCG sequences, as 
well as to possibly minimize its influence by penalizing large values of the
exchange splitting in the numerical search.  It is thus our hope that
significantly improved single-gate fidelities will be achievable in
S-T$_0$ qubits by aDCG sequences that operate under realistic
noise levels and control limitations.

\vfill
\section*{Acknowledgements}

It is a pleasure to thank Michael Biercuk, Matthew Grace, Robert Kosut, and
Amir Yacoby for valuable discussions and input.  Work at Dartmouth was supported by the
U.S. ARO (W911NF-11-1-0068), the U.S. NSF (PHY-0903727), and the IARPA
QCS program (RC051-S4).  HB was supported by the Alfried Krupp Prize
for Young University Teachers of the Alfried Krupp von Bohlen and
Halbach Foundation.


\begin{thebibliography}{100}
\bibitem[{bopt()}]{boptim} \bibinfo{note}{Changing the choice of the
drift may change the solutions considerably, for instance, having
$B=0$ removes controllability altogether. Note also that instead of
fixing $B$, we could in principle consider optimizing its value to be
fixed across a set of desired gates. }
\end{thebibliography}
\end{document}